\documentclass{optica-article}

\journal{opticajournal} 

\articletype{Research Article}

\newcommand{\Var}{\mathop{\mathrm{Var}} \nolimits}
\newcommand{\openone}{\leavevmode\hbox{\small1\normalsize\kern-.33em1}}
\newcommand{\Tr}{\mathop{\mathrm{Tr}} \nolimits}
\newcommand{\LG}[2]{\mathop{\mathrm{LG}}_{#1 #2} \nolimits}

\usepackage{graphicx,amsmath,relsize,epstopdf,color,mathtools,bm,newtxtext,newtxmath,braket,rotating}
\usepackage{lineno}

\begin{document}
\title{Enhancing axial localization with wavefront control}

\author{M. Peterek,\authormark{1} M. Pa\'ur,\authormark{1} M. V\'{\i}tek,\authormark{1} D. Koutn\'y,\authormark{1} B. Stoklasa,\authormark{1} L. Motka,\authormark{1} Z. Hradil,\authormark{1}  J. Rehacek,\authormark{1} L. L. S\'{a}nchez-Soto,\authormark{2,3,*}}

\address{\authormark{1}Department of Optics, Palack\'y University, 17.listopadu 1192/12, 779 00 Olomouc, Czech Republic\\
\authormark{2}Departamento de \'Optica, Facultad de F\'{\i}sica, Universidad Complutense, 28040 Madrid, Spain\\
\authormark{3}Max-Planck-Institut für die Physik des Lichts, 91058 Erlangen, Germany}

\email{\authormark{*}lsanchez@ucm.es}

\begin{abstract} 
Enhancing the ability to resolve axial details is crucial in three-dimensional optical imaging. We provide experimental evidence showcasing the ultimate precision achievable in axial localization using vortex beams. For Laguerre-Gauss (LG) beams, this remarkable limit can be attained with just a single intensity scan. This proof-of-principle demonstrates that microscopy techniques based on LG vortex beams can potentially benefit from the introduced quantum-inspired superresolution protocol.
\end{abstract}

\section{Introduction}
\label{sec:level1}

The Abbe-Rayleigh criterion~\cite{Abbe:1873aa,Rayleigh:1879aa} defines the minimum resolvable distance between two closely spaced point {sources in an image}. It is widely recognized, though, that the resulting limit is  inadequate for current quantitative imaging~\cite{Ram:2006aa}. In recent {years}, a collection of methods, which can be gathered under the rough denomination of superresolution microscopy~\cite{Hell:2007aa,Huang:2009aa,Schermelleh:2010aa,Leung:2011aa,Huszka:2019aa,Schermelleh:2019aa,Prakash:2022aa}, has emerged,  surpassing the length scale set by the Abbe-Rayleigh criterion by more than one order of magnitude.  

Among these remarkable methods are stimulated-emission-depletion microscopy~\cite{Hell:1994aa,Klar:1999aa}, stochastic optical reconstruction microscopy\cite{Rust:2006aa}, photoactivated-localization microscopy~\cite{Betzig:2006aa,Hess:2006aa}, structured illumination microscopy~\cite{Gustafsson:2008aa,Hirvonen:2009aa}, PSF engineering~\cite{Huang:2008aa,Pavani:2009aa,Jia:2014aa,Tamburini:2006aa,Paur:2018aa}, and multiplane detection~\cite{Juette:2008aa,Dalgarno:2010aa,Abrahamsson:2012aa}. {Notably, all of them} rely on the possibility of a {precise} localization of point sources.

In the broader context of three-dimensional microscopy, accurately determining the position of fluorescent objects along the optical axis is crucial~\cite{Diezmann:2017aa}. However, the challenge of finding the minimum resolvable separation in the axial direction has only {lately} been addressed~\cite{Tsang:2015aa,Backlund:2018aa}. The key idea is to resort to the quantum Fisher information~\cite{Helstrom:1976aa,Braunstein:1994aa} and the associated Cram\'er-Rao bound~\cite{Helstrom:1967aa,Belavkin:1976aa,Holevo:2003fv,Albarelli:2020aa} to get a measurement-independent limit. This builds upon the pioneering work of Tsang and coworkers to quantify transverse two-point resolution~\cite{Tsang:2016aa,Nair:2016aa,Ang:2016aa,Tsang:2017aa}.

Recently, a mode sorter capable of losslessly projecting photons into the radial LG basis set demonstrated quantum-limited estimation of axial separation for incoherent point sources~\cite{Zhou:2019aa}. This strategy bears similarities to techniques used in transverse superresolution~\cite{Paur:2016aa,Yang:2016aa,Tham:2017aa}. {Yet} the complexity of experimental setups, introducing systematic errors and losses, can undermine the theoretical advantages offered by {this optimal scheme.}

Surprisingly, it has been shown~\cite{Rehacek:2019aa} that this ultimate limit can be easily achieved with a single intensity scan when the detector is placed at one of two optimal transverse detection planes. This simplicity and feasibility make it highly valuable for applications requiring extremely precise localization. Additionally, this method has been extended to include vortex-beam illumination~\cite{Koutny:2021aa}. The goal of this paper is to experimentally demonstrate the localization limits in this particular scenario.

This paper is organized as follows. In Sec.~\ref{sec:limits} we revisit the theoretical framework needed to set the ultimate limits in axial resolution and how to achieve such limits. This ideal scenario is complemented in Sec.~\ref{sec:detrim} by taking into consideration detrimental effects, such as finite pixel size or misalignments, which are usually omitted. In Sec.~\ref{sec:exp} we present the experimental setup used to control the LG beam wavefronts using a spatial light modulator (SLM). In this way, we demonstrate the information gain provided by vortex beams and the experimental validation of the saturation of the quantum Cram\'er-Rao bound by intensity detection. Finally, our conclusions are summarized in Sec.~\ref{sec:conc}.

\section{Ultimate limits for axial localization}
\label{sec:limits}

In our quest to unravel the ultimate limits of axial resolution, we approach the challenge as the estimation of the distance $z$ traveled by a vortex beam from its beam waist, situated at $z=0$, to an arbitrary detection plane located at $z$. At this detection plane, we perform an arbitrary {measurement. The acquired data are always affected by noise, so they are effectively represented by a stochastic variable denoted by $\mathbf{x}$, enabling us to construct a robust estimator $\hat{z}$ for the distance $z$. The inference of the parameter $z$ is related to the measurement outcomes through some conditional probability density $p(\mathbf{x} | z)$ that is dictated by the process at hand.}

{To quantify the maximum information carried by the measured signal, we turn to the concept of Fisher information (FI), defined as}
\begin{equation}
\label{eq:cfi}
{\mathcal{F} (z) = \int 
\left [ \partial_z  \ln p(\mathbf{x}|z) \right]^{2} p(\mathbf{x}|z) \, d\mathbf{x} = \int 
\frac{\left[\partial_z p(\mathbf{x}|z)\right]^2}{p(\mathbf{x}|z)} \,  d\mathbf{x} \, .} 
\end{equation}

{In the quantum domain we use a probe state given by the density operator $\varrho$. The propagation encodes the parameter $z$ via the transformation $\varrho_{z} = U_{z} \varrho U_{z}^{\dagger}$, with the unitary operator $U_{z} = \exp (- i z G)$, where $G$ is the generator of translations. The measurement is represented by some positive operator-valued measure (POVM) $\{ \Pi_{\mathbf{x}} \}$, which comprises a set of positive semidefinite selfadjoint operators that resolve the identity~\cite{Holevo:2003fv}. By performing this measurement, we obtain a statistical distribution that, according to Born's rule, is given by $p(\mathbf{x} | z) = \Tr ( \varrho_{z} \,  \Pi_{\mathbf{x}})$.  Afterward, what remains is to obtain the best estimate of $z$ given  $p(\mathbf{x} | z)$.}

{It is natural to ask whether there is an optimal measurement that should be performed on $\varrho_{z}$. The quantum Fisher information (QFI) is defined precisely for this purpose: $\mathcal{Q}_{\varrho} (z ) = \sup_{ \{ {\Pi}_{\mathbf{x}} \} } \mathcal{F} (z)$ and depends exclusively on the probe state $\varrho$. By its very same definition, we have $\mathcal{Q}_{\varrho} ( z ) \geq \mathcal{F} ( z)$.}

{According to the time-honored quantum Cram\'er-Rao bound (QCRB), the variance of any unbiased estimator $\hat{z}$ per detection event of the axial distance $z$ is bounded by}
\begin{equation}
\label{eq:QCRB}
\Var_{\varrho} (\hat{z}) \geq \frac{1}{\mathcal{Q}_{\varrho} (z)} \, .
\end{equation}
The right hand side of Eq.~(\ref{eq:QCRB}) thus represents the ultimate achievable precision regardless of the measurement {and depends exclusively on  the beam used in the experiment.}  For definiteness, we take the beam to be in an $\LG{p}{\ell}$ mode, with complex amplitude given by~\cite{Yao:2011aa} 
\begin{align}
\label{eq:LGDef}
\LG{p}{\ell}(r,\phi,z)  & =  \sqrt{\frac{2p!}{\pi (p+|\ell|)!}} \frac{w_{0}}{w(z)}\left [\frac{\sqrt{2}r}{w(z)}\right]^{|\ell|}  \,  \exp \left  [ - \frac{r^2}{w(z)^2} \right ]  & \nonumber \\
& \times L_p^{|\ell|} \left(\frac{2r^2}{w(z)^2}\right)  
 \exp \left( i \left [ \frac{kr^{2}}{2 R(z)} - \ell \phi - \psi_{p \ell}(z) \right ] \right) \, ,  &   
\end{align}
where $(r, {\phi},z)$ are cylindrical coordinates, $L_{p \ell} (\cdot)$ is a generalized Laguerre polynomial~\cite{NIST} with radial index {$p \in \{ 0, 1, 2, \ldots \}$ and azimuthal mode index $\ell \in \{ 0, \pm 1, \pm 2, \ldots \}$}. In this expression, the following parameters
\begin{equation}
   \begin{split}
       R(z) & =z \left[1+(z_{\mathrm{R}}/z)^2\right],\\
       w(z)^{2} & =w_0 \left[1+(z/z_{\mathrm{R}})^2\right],\\
    \psi_{p\ell}(z)& =(2p+|\ell| +1) \,  \arctan(z/z_{\mathrm{R}} ),\\
   \end{split}
   \label{eq:2}
\end{equation}
\noindent
are the radius of wavefront curvature, the beam radius, and the Gouy phase, respectively, at a distance $z$ from the beam waist.  {Here}, $z_{\mathrm{R}} = k w_0^2/2$ is the Rayleigh length and $w_{0}$ the beam waist radius~\cite{Siegman:1986aa}. 

The QFI for an arbitrary $\LG{p}{\ell}$ mode has been recently worked out in Ref.~\cite{Koutny:2021aa}; the result reads
\begin{equation}
    \mathcal{Q}_{p\ell} (z)=\frac{1}{z_R^2}\left[2p(p+|\ell|)+2 p+|\ell|+1\right] \, ,
    \label{eq:4}
\end{equation}
which reduces to  $\mathcal{Q}_{00}(z)=1/z_R^2$ for the fundamental Gaussian mode $\LG{0}{0}$. Note that {this} QFI is linear in $\ell$, and so the axial localization can be  significantly improved by using higher LG modes.

\begin{figure}[t]
\centering \includegraphics[height=5.5cm]{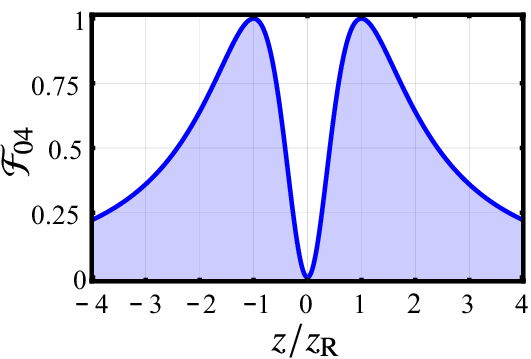}
\caption{{Fisher information} for direct intensity detection with the mode $\LG{0}{4}$ as a function of the axial coordinate $z$. At   $z=\pm z_{\mathrm{R}}$,  the function $\mathcal{F}_{04}$ reaches its maximal value; these are the optimal planes to place the detector.}
\label{fig:1}
\end{figure}

Since we are dealing with a single-parameter estimation, the QCRB can always be saturated with a von Neumann measurement projecting the measured signal on the eigenstates of the symmetric logarithmic derivative of the density matrix~\cite{Helstrom:1976aa}. This implies projecting the signal onto a complete set of optimal modes, which requires a fairly sophisticated and fragile equipment. Therefore, we consider the performance of the possibly inferior, but much more robust scheme of direct intensity detection, as this is the handiest method available in the laboratory. 

{We assume that the intensity is sampled by a pixelated detector and the signal is dominated by shot noise, which obeys a Poisson distribution~\cite{Fuhrmann:2004aa}: this neglects nonclassical effects,  but it is still a suitable model for realistic microscopy. For the time being we ignore the pixel size so any sampling effect can be omitted. We can take $p(r,\phi|z) = |\LG{p}{\ell}(r, \phi, z)|^{2}$, which has to be understood as the probability that we observe a given transverse intensity at $(r, \phi)$, given a known value of $z$.} Then, Eq.~\eqref{eq:cfi} admits a closed expression; after a lengthy calculation one gets~\cite{Koutny:2021aa}
\begin{equation}
\label{fish_intens}
{\mathcal{F}_{p\ell} (z) =  \int_0^{\infty}\int_0^{2\pi} 
\frac{\left[\partial_z p(r,\phi|z)\right]^2}{p(r,\phi|z)} \, 
r drd\phi = \frac{2p(p+|\ell|)+2p+|\ell|+1}{ \tfrac{1}{4} R(z)^{2}} \, .}
\end{equation}
This function is plotted in Fig.~\ref{fig:1} for the beam $\LG{0}{4}$.  We appreciate the existence of two well-defined maxima located precisely at $z_{\mathrm{opt}}= \pm z_{{\mathrm{R}}}$. {Actually, maxima occur at $z_{\mathrm{opt}}= \pm z_{{\mathrm{R}}}$ for any $\LG{p}{\ell}$ mode}.  In these two planes, we have maximal wavefront curvature and one can check that {$\mathcal{Q}_{p\ell} (z_{\mathrm{R}}) = \mathcal{F}_{p\ell}  (\pm z_{\mathrm{R}})$}; i.e., the quantum limit is saturated. Therefore, in these two detection planes complete information about the axial distance can be extracted with intensity-only detection. 

\begin{figure}[t]
\centering \includegraphics[height=5.5cm]{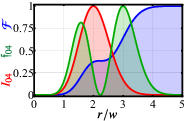}
\caption{Plots of the cross-section of the intensity distribution $I_{04}$ (red), the FID $\mathfrak{f}_{04}$ (green) and the cumulative FI $\mathcal{F}$ (blue) for a pure $\LG{0}{4}$ beam, as a function of the dimensionless transverse distance $r/w$ at the optimal detection plane $z_{\mathrm{R}}$.  Each curve is normalized to unity.}
\label{fig:2}
\end{figure}

\section{Detrimental effects}
\label{sec:detrim}

In the ideal scenario discussed earlier, certain imperfections and realistic factors were omitted, which can undermine the anticipated metrological advantages. It is crucial to acknowledge and account for these considerations in any practical setup.

A noteworthy observation is that the integrand in Eq.~\eqref{fish_intens} can be aptly interpreted as a Fisher information density (FID), we shall denote by $\mathfrak{f} (r, \phi | z)$. This FID reveals the data with highest sensitivity to variations in the coordinate $z$ and is  plotted in Fig.~\ref{fig:2} for the particular case of a $\LG{0}{4}$ beam. An intriguing feature is the presence of two highly sensitive regions: the inner and outer tails of the beam, with the latter being particularly informative. We thus conclude that there is no necessity to gather data from an area larger than 5 times the beam radius, as it contributes no valuable information for axial localization. In fact, collecting data beyond this range would only introduce additional noise, further compromising the precision.

%
\begin{figure}[t]
\centering\includegraphics[height=5.5cm]{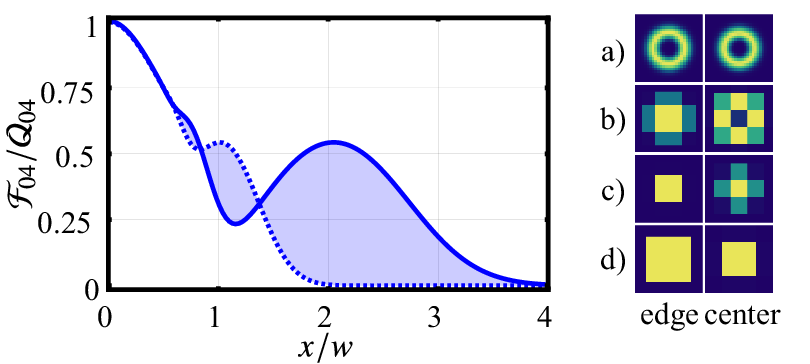}
\caption{{(Left) Fisher information in direct detection at the optimal plane for the beam $\LG{0}{4}$ (normalized to the optimal value $\mathcal{Q}_{04}$) as a function of the pixel size (normalized as $x/w$).  For the full line the integration starts at the pixel center, whereas for the dotted line the integration starts at the edge of the individual pixel. (Right)  Integrated signal for both strategies for the pixel sizes: a) $0.5x/w$, b) $1.5 x/w$, c) $2.5x/w$, d) $3.5x/w$.}}
\label{fig:3}
\end{figure}
\begin{figure}[b]
\centering\includegraphics[height=5.5cm]{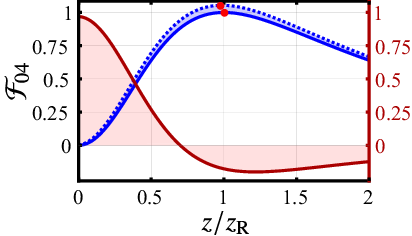} 
\caption{{The blue continuous curve represents the Fisher information for intensity detection with the mode $\LG{0}{4}$ as a function of the axial coordinate $z/z_{\mathrm{R}}$, as in Fig.~\ref{fig:1}. The blue dotted curve is the Fisher information for the same mode, but with a lateral centroid deviation of 10~mrad. The red dots denote the maxima of both curves, and thus the position of the corresponding optimal detection planes. For the misaligned beam, this optimal plane shifts toward the beam waist (located at $z=0$). The brown curve is the normalized differential gradient of the Fisher information with and without misalignment.}}
\label{fig:4}
\end{figure}
Another point to take into account is the finite pixel size. We have to interpret each pixel as an independent measurement channel, so that
\begin{equation}
p(\textbf{x}|z)=p(x_1,x_2,\ldots  ,x_n|z)=\prod_{j=1}^n p_j(r_{j}, \phi_{j}|z),
\label{eg:7}
\end{equation}
where $p_j(r_j,\phi_j|z)=|\LG{p}{\ell} (r_j,\phi_j|z)|^2$  is the conditional probability for the $j$-th measurement channel.  We numerically examine this effect. Intuitively, one might anticipate a steady decrease in information as the pixel size increases. This is attributed to the integration of the signal across individual pixels, causing the gradients in Eq.~\eqref{fish_intens} to blur. 

However, as revealed in Fig.~\ref{fig:3}, the situation is not universally characterized by such a monotonic decrease. Remarkably, when considering a $\LG{0}{4}$ beam, we observe that using a pixel size 3 times the beam radius actually yields more information than using a pixel size 2 times the beam radius. This counterintuitive phenomenon arises due to specific pixel sizes that integrate the signal from pixels with low FID, while the edge of the adjacent pixel is close or exactly in the area with highest gradient, with the highest FID. Consequently, the complete LG beam is integrated into a single pixel with significantly large pixel sizes, resulting in a loss of information. It is important to emphasize that this behavior is strongly dependent on the sampling mesh grid employed.
 
Another critical issue in the experiment is the misalignment of the mechanical and optical axes. Due to the different propagation angles, a lateral centroid movement is measurable by the detector. The simple model we use here is based on straight light propagation:
\begin{equation}
    \Delta x(z) = z \tan \varphi \, ,
    \label{eg:9}
\end{equation}
where $\Delta x$ is lateral centroid motion at the detection plane $z$ and $\varphi$ is angle between the optical and mechanical axis. Because $\Delta x=\Delta x(z)$ depends on the estimated parameter, lateral centroid motion becomes an additional information source in the axial localization problem. This is illustrated in Fig.~\ref{fig:4}, where we have also included the gradient of the FI variation, which turns important to assess the dominant information source: positive values give lateral centroid motion, while negative values are mainly due to diffraction.


\section{Experiment}
\label{sec:exp}

\begin{figure}[b]
\centering \includegraphics[width=\columnwidth]{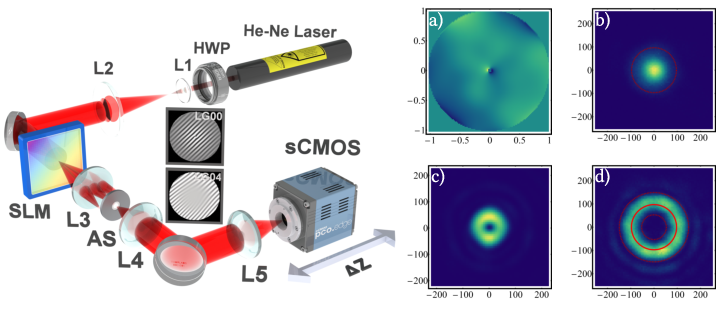}
\caption{\label{fig:5} {(Left) Scheme of our experimental setup. (Right) a) SLM wavefront used for the GS algorithm. The measured corrected beams appear in b) $\LG{0}{0}$, c) $\LG{0}{1}$, and d) $\LG{0}{4}$. The red circles indicate the radii for maximum intensity and corresponding inner and outer radii with intensity drop to $1/e^2$.}}
\end{figure}

Our experimental setup is schematized in Fig.~\ref{fig:5}. A He-Ne laser at wavelength $\lambda=$~633~nm is spatially filtered by {an} aperture stop (AS) and {expanded and collimated} by two doublet lenses, $L_{1}$ and $L_{2}$. {This} collimated beam illuminates the SLM (CRL OPTO XGA3) with 18~$\mu$m-pixel size.  A  computer-generated hologram (CGH) is imprinted on the SLM to generate LG vortex beams with a given azimuthal mode index. The {pixelated nature} of the SLM leads to multiple diffraction orders, which appear in the image plane of the lens $L_{3}$. To separate the diffraction orders, we mixed {the} CGH with a plane wave and filter using a $4f$ system with an AS. We also effectively relocate the conjugate plane from SLM to a lens $L_{5}$, the true imaging lens in the experiment. The $L_5$ has a focal length of 150~mm, and the CGH is 2~mm in diameter. 

The measured intensity is proportional to $|\LG{p}{\ell}|^2$. A compact sCMOS camera (pco.edge 4.2 LT) with a very high dynamic range ({up to $37500:1$}) was used. {The detection noise (assumed as Poissonian) is about 100 electrons (with a standard deviation of 8 electrons), so it is negligible. In addition, the saturation level is of the order of 30000 electrons, which provides a fairly good signal-to-noise ratio.}  

{The beam was characterized with a standard Beam Analyzer software. The analysis is based on interpolating the beam profile and measuring the width at $1/e^{2}$ for several (in our case, eight)  cross sections. The Rayleigh range is calculated from the measured beam waist. The beam width for $\LG{0}{0}$ was compared with the theory at several planes with good match.}

We first calibrated the SLM nonlinear response, getting an input-output relation by using several uniform amplitudes. Next, we employed the well-established Gerchberg-Saxton (GS) iterative algorithm~\cite{Gerchberg:1972aa} to correct the SLM wavefront error. With just one hundred iterations, guided by an initial estimation of the typical astigmatism shape, we successfully restored the wavefront to a satisfactory state. To further refine the system, we employed a modal decomposition in terms of  Zernike polynomials~\cite{Love:1997aa}, {minimizing the residual wavefront errors.} The results can be observed in Fig.~\ref{fig:5}.

We observed a deviation in the SLM correction when dealing with azimuthal mode indexes other than $\ell=1$. This discrepancy was attributed to a systematic error arising from changes in the propagation angle on the SLM with varying $\ell$ values. To mitigate these errors, we developed a customized data processing, ensuring optimal performance and accuracy.

{To check the behavior of the QCRB near the Rayleigh range $z_{\mathrm{R}}$}  we suppose small displacements $\delta z$ around this plane $z=z_{\mathrm{R}}\pm \delta z$, with  $\delta z \ll  z_{\mathrm{R}}$. We manually moved by $\delta z$ the sCMOS camera using a linear stage (STANDA 7T167-100Q) with a positioning precision of 5~$\mu$m.  We collected the data stack at five detection planes located at  $\delta z_\alpha= -200~\mu$m, $- 100~\mu$m, $0~\mu$m,  $ 100~\mu$m, and $200~\mu$m {from $z_{\mathrm{R}}$.}  Each stack is comprised of 100 intensity frames with a total number of detections per frame  about $N=7\times 10^5$ events  in average (with background subtracted). The same procedure was repeated for other $\LG{p}{\ell}$ beams up to the azimuthal order $\ell=7$, which was the last measurable beam in our setup.  In addition, the lateral centroid motion was subtracted from the data.

In the range of measured $\delta z$ values, the changes in intensity are small. This suggests to adopt a polynomial basis so that the average intensity $\bar{\mathcal{I}}_j(\delta z)$ at the {$j$th} pixel can be expressed as
\begin{equation}
\label{model}
\bar{\mathcal{I}}_j ( \delta  z) = \sum_k  C_{kj} \,\delta z^k \, ,
\end{equation}
where $C_{jk}$ is the model coefficient matrix. To estimate this matrix we use the five data sets at planes $\delta z_{\alpha}$. Let us denote by $\bar{\bm{\mathcal{I}}}$ and $\bm{\delta z}$ the matrices of components $\bar{\mathcal{I}}_{\alpha j} = \bar{\mathcal{I}}_j ( \delta z_{\alpha} ) $ and  $\delta z_{\alpha k}= \delta z_{\alpha}^{k}$, respectively. We first solve for the model coefficient matrix using the linear inversion estimator:
\begin{equation}
\hat{\mathbf{C}} = \bm{\delta z}^{(+)} \; \bar{\bm{\mathcal{I}}} \, ,
\end{equation}
where the superscript $(+)$ indicates the Moore-Penrose {pseudoinverse}~\cite{Ben-Israel:1977aa,Campbell:1991aa}. This estimator is also known as the ordinary least squares  estimator~\cite{Lawson:1974aa} and it is unbiased and consistent.  Under the Gauss-Markov assumptions it is also the best linear unbiased estimator~\cite{Hallin:2006aa}.

Once $\hat{\mathbf{C}}$ is known, we can estimate the axial displacement from the set of frames recorded at the Rayleigh plane, denoted as $\bar{I}_j^{(\beta)}$,  where $\beta = 1, \ldots, 100$. Now, the relation reads 
 \begin{equation}
\label{planes}
\bar{\mathcal{I}}_j^{\beta} = \sum_k  C_{kj} \,\delta z^{(\beta) \,k} \, .
\end{equation}
This relation can be inverted by the generalized linear squares method~\cite{Lawson:1974aa} and  finally, the statistics of displacement estimates $\delta \hat{z}^{(\beta)}$ is evaluated and FI {per single detection event} estimated by taking 
\begin{equation}
\mathcal{F} \simeq \frac{1}{N \Var (\delta \hat{z})} \, .
\end{equation}

\begin{figure}[t]
\centering\includegraphics[height=5.5cm]{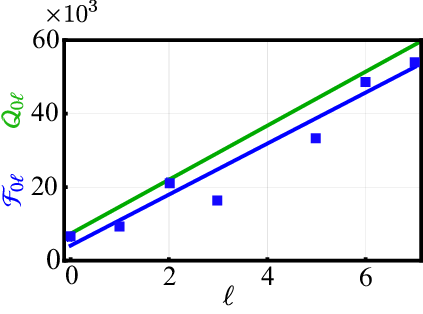} 
\caption{Green line depicts the QFI as a function of the azimuthal mode index $\ell$ of a pure $\LG{0}{\ell}$ mode. The blue dots are the measured FI for direct detection, as explained in the text. The blue line is a linear fit of the points.}
\label{fig:6}
\end{figure}

In Fig.~\ref{fig:6} we represent the resulting {FI} computed from our experimental data as a function of the azimuthal mode index $\ell$. For comparison, we also plot the QFI. We also include a linear fit of the experimental points that confirms the linear growth of {the  FI with $\ell$.} This line although quite parallel to the QFI, is a bit below, {because of the}  abnormally low value of the FI for $\ell = 3$, due to the experimental difficulties. The rest of the points are close enough to the ultimate limit. Interestingly, we see an apparent violation of the QFI limit for $\ell = 4$, but this is surely due to systematic errors. 

{The results in this figure confirm that our protocol is able to determine the axial distance to an accuracy grater than the classical predictions based in the Abbe-Rayleigh approach, which report a depth of focus of the order of the Rayleigh range. If we take, e.g., the data from the $\LG{0}{7}$  beam, we get a standard deviation of $0.000432455$~m per detection. With $N = 7 \times 10^{5} $ detections registered, this gives $5.1688~\mu$m, beating the classical limit ($11.67$~mm) by three order of magnitude. Similar estimates can be carried out for any LG beam.}

\section{Concluding remarks}
\label{sec:conc}

In summary, we have unveiled the extraordinary potential of pure LG vortex beams for achieving unparalleled precision in axial localization. Through an extensive experimental analysis, we have uncovered the ultimate limits and demonstrated the remarkable advantages of these beams. Our findings reveal an exciting trend: as the azimuthal mode index increases, so does the precision, unlocking a world of possibilities. However, it is crucial to underline the inherent experimental challenges in controlling these vortex-beam wavefronts with SLMs.

Harnessing the power of FI, we have delved into the realm of detrimental effects that have been overlooked in the existing literature. Finite pixel size and the misalignment of optical and mechanical axes, which are often dismissed, have now been brought to the forefront. 

{Our results offer new insights into the localization problem and open up new avenues for exploration that might be interesting for 3D microscopy.}

\begin{backmatter}
\bmsection{Funding}
European Union’s Horizon 2020 Research and Innovation Programme (QuantERA ApresSF);  
Palack\'y University (IGA$\_$PrF$\_$2023$\_$002); 
Spanish Ministerio de Ciencia e Innovaci\'on (PID2021-127781NB-I00).

\bmsection{Acknowledgments}
We ackowledge discussions with M. Sondermann. 

\bmsection{Disclosures}
The authors declare no conflicts of interest.

\bmsection{Data availability} 
Data underlying the results presented in this paper are not publicly available at this time but may be obtained from the authors upon reasonable request.

\end{backmatter}


\end{document}